\documentclass[conference]{IEEEtran}
\IEEEoverridecommandlockouts
\usepackage{cite}
\usepackage{amsmath,amssymb,amsfonts}
\usepackage{algorithmic}
\usepackage{algorithm}
\usepackage{graphicx}
\usepackage{textcomp}
\usepackage{xcolor}
\def\BibTeX{{\rm B\kern-.05em{\sc i\kern-.025em b}\kern-.08em
    T\kern-.1667em\lower.7ex\hbox{E}\kern-.125emX}}
\begin{document}

\title{Generalization of higher order methods for fast iterative matrix inversion via GPU acceleration\\

}

\author{\IEEEauthorblockN{1\textsuperscript{st} Marcus Engsig}
\IEEEauthorblockA{\textit{Directed Energy Research Centre} \\
\textit{Technology Innovation Institute}\\
Abu Dhabi, UAE \\
marcus.engsig@tii.ae}
\and
\IEEEauthorblockN{2\textsuperscript{nd} Qingjie Yang}
\IEEEauthorblockA{\textit{Directed Energy Research Centre} \\
\textit{Technology Innovation Institute}\\
Abu Dhabi, UAE \\
qingjie.yang@tii.ae}
\and
\IEEEauthorblockN{3\textsuperscript{rd} Fares Mehouachi}
fares.mehouachi@gmail.com}

\maketitle

\begin{abstract}
Recent technological developments have led to big data processing, which resulted in significant computational difficulties when solving large-scale linear systems or inverting matrices. As a result, fast approximate iterative matrix inversion methodologies via Graphical Processing Unit (GPU) acceleration have been a subject of extensive research, to find solutions where classic and direct inversion are too expensive to conduct. Some currently used methods are Neumann Series (NS), Newton iteration (NI), Chebyshev Iteration (CI), and Successive Over-Relaxation, to cite a few. In this work, we develop a new iterative algorithm based off the NS, which we named 'Nested Neumann' (NN). This new methodology generalizes higher orders of the NI (or CI), by taking advantage of a computationally free iterative update of the preconditioning matrix as a function of a given 'inception depth'. It has been mathematically demonstrated that the NN: (i) convergences given the preconditioning satisfies the spectral norm condition of the NS, (ii) has an order of rate of convergence has been shown to be equivalent to the order (inception depth plus one), and (iii) has an optimal inception depth is an inception depth of one or preferably two, depending on RAM constraints. Furthermore, we derive an explicit formula for the NN, which is applicable to massive sparse matrices, given an increase in computational cost. Importantly, the NN finds an analytic equivalancy statement between the NS and the the NN (NI, CI, and higher orders), which is of importance for massive multiple input multiple output systems, as assessment accuracy of the inverse using these methods can be analytically compared. Finally, the NN method is applicable positive semi-definite matrices for matrix inversion, and applicable to any linear system (sparse, non-sparse, complex, etc.).
\end{abstract}
\vspace{2mm}


\section{Introduction}

During the Big Data revolution witnessed in the last few decades, the complexity of linear systems, machine learning, and data processing algorithms has increased to match the current processing capabilities. Linear system resolution or matrix inversion is an essential linear algebra operation that often appears in various modeling, and engineering problems \cite{Hashima2020, Oscar2017, Zhu2015, Shao2016, Albreem2021OverviewOP, Gil2018, moulinec2018convergence, qian2011efficient,  lorraine2020optimizing, krishnan2017neumann, fung2022jfb}. In this work, we focus primarily on matrix inversion and linear system resolution. Aside from exceptional cases (eg. identity matrix), direct matrix inversion methods scale with the cube of the dimension of the problem (ie. $O(N^3)$). As a result, despite technological advancements, large-scale inversion tasks remain challenging and computationally expensive. Therefore, interest in iterative methods for fast matrix inversion has grown steadily during the last two decades \cite{Shao2018ApproachesOA, haghani2014new, pan1991improved}. Such methods commonly avoid direct inversion and focus on using embarrassingly parallelizable operations (ie. matrix multiplication, addition, etc.) via Graphical Processing Units (GPUs). With Moore's law reaching a plateau \cite{moore} these last few years, parallelization singularly remains the viable path to tackle high-dimensional inversion problems. Fast iterative approximate matrix inversion algorithms have been studied extensively, and some libraries exist and are open source, such as cublas \cite{cublas}, cusparse, and AmgX \cite{amgx}. These libraries provide a multitude of ready-to-use methods, yet matrix inversion remains challenging, and there is still potential room for improvement. This work presents a new method, called the \textit{Nested Neumann} (NN), which is applicable to massive sparse systems, and small dense systems, yet needs empirical performance testing with a powerful GPU and optimized systolic arrays, in line with Pan and Schreiber's suggestions \cite{pan1991improved}.

\section{Methodology}
Fast matrix inversion algorithms are based on iterative methods that use exclusively embarrassingly parallelizeable operations, like the \textit{Neumann Series} (NS), \textit{Newton} (NI), or \textit{Chebyshev Iterations} (CI) \cite{Hashima2020, Oscar2017, Zhu2015}. These methods converge to the solution with an increasing number of iterations, implying a trade-off between the computational time ($\tau$) and the mismatch error ($\epsilon$: root mean Frobenius norm error) with the real solution. These methods generally require or are used for generating a computationally efficient and suitable preconditioning matrix. The preconditioning matrix has two functions: (i) to normalize the spectral radius of the matrix product to ensure convergence, and (ii) to increase the speed of convergence. Generally, the quality of the preconditioning matrix increases as it approaches the inverse, which determines the number of steps required for the iterative methods to converge.
\subsection{Neumann Series (NS)}
Since the seminal work by Carl Neumann, a majority of the fast iterative approximate matrix inversion methodologies revolve around the NS \cite{Hashima2020, Oscar2017, Zhu2015}. The NS for inverting a positive semi-definite matrix $W \in \mathbb{C}_{N\times N}$, with a preconditioning matrix $\phi$ can be expressed as;
\begin{equation}
\label{eq:Neumann}
  W^{-1} \approx \sum_{n=0}^{L} [(I - \phi W)^n] \phi, \quad ||I - \phi W ||_2 < 1 ,
\end{equation}
where $L$ is the Neumann iteration number, which provides a tradeoff between accuracy and computational cost, $I \in \mathbb{C}_{N\times N}$ is the identity matrix of dimension N, and $||I-\phi W||_2 <1$ is the necessary spectral norm condition for convergence. The preconditioning matrix can be used to normalize this matrix product. For practicality, we define the normalized matrix $\tilde{W}$ that satisfies the spectral norm condition: 
\begin{equation}
    \tilde{W} = \Theta W : ||I-\tilde{W}||_2 < 1,
    \label{eq:normalization}
\end{equation}
where $\Theta$ satisfies this criteria for positive and semi-definite (PSD) matrices if $\Theta \in \{\Theta_1, \Theta_2\}$,
\begin{equation}
\Theta_1 = \frac{1}{\text{Tr}(W)}, \quad \Theta_2(k) = \frac{k||W^k\boldsymbol{v}||^2_2}{(k+1)||W^{k+1}\boldsymbol{v}||^2_2}
\label{eq:trace}
\end{equation}
where $\boldsymbol{v}$ is a random vector, and the order $k$ determines the assessment accuracy of the largest eigenvalue estimation (for acceleration use $k=2^n, n\in\mathbb{N}$). Here we found, that the worse the conditioning of the matrix, the less benefit there is from using $\Theta_2(k)$, as $\Theta_1 \to \Theta_2$ when one eigenvector becomes increasingly dominant.

\subsection{Newton Iteration (NI)}
The NI is a matrix analogy to finding a minima or zero using the Newton second order approximation, which has a quadratic rate of convergence. The NI for functions is defined:
\begin{equation}
    Z^{(n+1)} = Z^{(n)} - \frac{f(Z^{(n)})}{f'(Z^{(n)})},
    \label{eq:newtonfunction}
\end{equation}
which can be transferred to finding the inverse of a matrix by substituting $f(Z) = Z_2^{-1} - \tilde{W}$, which can be solved to yield the following equation  \cite{Hashima2020, Shao2018ApproachesOA}, given that $|| I - \tilde{W} ||_2 < 1 $;
\begin{equation}
    Z^{(n+1)} = Z^{(n)}(2I - W\tilde{Z}^{(n)}),
    \label{eq:newtoniteration}
\end{equation}
where $Z^{(n)}$ converges towards the inverse of $\tilde{W}$, as $n\to\infty$, given an initial guess $Z^{(0)}$ in the broad neighborhood of $\tilde{W}^{-1}$.

\subsection{Chebyshev Iteration} 
The Chebyshev Iteration (CI) is the third order variation of the NI, which has a cubic rate of convergence. Due to its high computational complexity, the CI is rarely used, and only appears frequently in providing an estimate for the preconditioning matrix for massive MIMO systems \cite{ Zhu2015}. The CI for function is displayed below,
\begin{equation}
    Z^{(n+1)} = Z^{(n)} - \frac{f(Z^{(n)})}{f'(Z^{(n)})} - \frac{f''(Z^{(n)})}{2f'(Z^{(n)})}\left( \frac{f(Z^{(n)})}{f'(Z^{(n)})} \right)^2,
    \label{eq:chebyshevfunction}
\end{equation}
that also converges to the inverse of a matrix, given the same normalization condition as for the Neumann Series \cite{Shao2018ApproachesOA}, that is $||I - \tilde{W}||_2 < 1$ . For matrices, the Chebyshev iteration can be solved similarly to the Newton iteration, and it can be stated as follows, given an initial guess $Z^{(0)}$:
\begin{equation}
\label{eq:cheby}
Z_3^{(n+1)} = Z^{(n)}[3I - \tilde{W}Z^{(n)}(3I - \tilde{W}Z^{(n)})].
\end{equation}

The work by Pan and Schreiber in 1991 \cite{pan1991improved}, studied the NI and the CI in depth, concluding that the computational complexity of these methods was too high, and attributed it would only be applicable to invert matrices with a number of cores comparable with the dimension of the matrix. Since then, few people have tried higher order methods \cite{haghani2014new}, and the NI and CI usually are only used for a few iterations to provide preconditioning matrices for the NS \cite{Shao2016}. Now, with the emergence of GPU clusters and powerful accessible GPUs, the hypothesis by Pan and Schreiber is becoming more testable, and perhaps realizable. 

\subsection{Nested Neumann}
In this work, we introduce a new method, the Nested Neumann (NN), which focuses on creating an efficient iterative update of the preconditioning matrix to increase the speed of convergence without inducing a high computational cost in the search of a suitable preconditioning matrix.\smallskip

\noindent \textbf{Theorem:}
Given a positive semi-definite matrix $W\in\mathbb{C}_{N\times N}$, we can approximate the inverse of $W$, by using an iteratively updated preconditioning matrix $\phi_{L}^{(i)}$, given that $W$ is normalized by $\tilde{W} = \Theta W$, such that $|| I - \Theta W ||_2 < 1$,
\begin{equation}
  W^{-1} \approx \phi_{L}^{(i)}\Theta,
\end{equation}
where $\phi_{L}^{(i)}$ is defined iteratively:
\begin{equation}
 \tilde{W}\approx \phi_{L}^{(i+1)} = \sum_{n=0}^{L}(I-\phi_{L}^{(i)}\tilde{W})^n\phi_{L}^{(i)},
 \label{eq:nestedneumann}
\end{equation}
where $i$ is the number of nests, and $L$ is the denoted inception depth. Fast convergence with a high order of accuracy can be achieved with relatively small $i,L\in\mathbb{N}$ (see \textit{corollary 5}), with an order of the rate of convergence of $L+1$ as shown in \textit{Appendix (5)}.

The expectation might be that the fastest convergence would occur when $L=1$, as the NN takes full advantage of the iterative update. Yet, in actuality, the fastest convergence occurs when $L=2$ (see \textit{Appendix (6)}), which is discussed in detail in Corollary 2. The initial guess for $\phi^{(0)}$, is any preconditioning matrix that satisfies $||I-\phi^{(0)}\tilde{W}||<1$, which, here is chosen to be $\phi^{(0)} = I_{N\times N}$ as it is computationally free, and yields fast convergence. Many different initial guesses have been tested, but the noticed trend is that the reduction in the number of required total nests from a better initial guess of $\phi^{(0)}$ is computationally speaking not worth its computational cost to generate (if guaranteed convergence is desired). For proof of convergence $\forall i,L\in \mathbb{N}$, see \textit{Appendix (2)}. We find that the computational cost $C$ of the NN is,
\begin{equation}
    C = i(L+1)N^3 + (i+1)(L+1)N^2,
    \label{eq:compcost}
\end{equation}
which is notably very high, but also extremely parallelizable. Thus, it can be implemented with high efficiency for systolic arrays on GPUs.  \cite{pan1991improved}. 
\smallskip

\noindent \textbf{Corollary 1:}
The NN can be written in a non-recursive manner, such that $\phi^{(i)}_L(i,L,\phi^{(0)})$ is a function of the number of nests, inception depth, and the initial guess (see \textit{Appendix (3)}),
\begin{equation}
     \phi_L^{(i+1)}(i,L,\phi^{(0)}, \tilde{W}) =  \prod_{j=0}^{i}\sum_{n=0}^{L}(I-\phi^{(0)} \tilde{W})^{n(L+1)^j}\phi^{(0)}.
     \label{eq:NNexplicit}
\end{equation} 
Here we see that the NN is a function of only the tuning parameters $i,L$ and the initial guess $\phi^{(0)}$. We also show that the computational cost of this method without storing any matrices (applicable for massive sparse matrices):
\begin{equation}
\begin{split}
    C = \frac{L}{2}[i(\log_{L+1}(\gamma+1)(\log_{L+1}(\gamma+1)+1)+1]N^3 \\ + [2i\log_{L+1}(\gamma+1)+1]N^3
    \label{eq:compcostfactorizedsparse}
\end{split}
\end{equation}
where $\gamma = (L+1)^{i+1} -1$, which makes the NN applicable to sparse matrices at the expense of computational cost. However, for certain orders, it is equivalently competitive with the original NN. 
\smallskip

\noindent \textbf{Corollary 2:}
The NN can be rewritten into a NS in order to see the effect of the $i,L$ parameters (see \textit{Appendix} (4)),

\begin{equation}
        \phi^{(i+1)} = \sum_{n=0}^{(L+1)^{i+1}-1}(I-\phi^{(0)}\tilde{W})^n \phi^{(0)}.
    \label{eq:NNasNS}
\end{equation}
Here, we see that the inception depth functions as the base of an exponential $(L+1)$, with the power being the number of nests $i$. Thus, given the computational complexity of the NN, as in eq. \ref{eq:compcost}, the optimal inception depth is easily found to be $L= e - 1$. In practicality, this means $L=1$ or $L=2$. Here, the benefit of $L=1$ is reduced RAM requirements (storage of $I-\phi^{(i)}\tilde{W}$ is not necessary), but at an increased computational cost compared to $L=2$ (ie. $2^3 < 3^2$). 
\smallskip

\noindent \textbf{Corollary 3:}
Given any matrix $A\in \mathbb{C}^{N\times N}$ and $B \in \mathbb{C}^{N\times k}$, the linear system:
\begin{equation}
    A\boldsymbol{x} = B,
    \label{eq:linearsystem}
\end{equation}
 can be solved with the Nested Neumann with the following algorithm. Firstly, multiplying by  the complex conjugate $A^*$ and solving for $\boldsymbol{x}$, eq. \ref{eq:linearsystem} yields,
 \begin{equation}
     \boldsymbol{x} = (A^*A)^{-1}A^*B,
     \label{eq:linearsystemfinal}
 \end{equation}
 and by letting $W = A^*A$, we know that by construction, $W$ is a PSD matrix. Thus, we can use the NN of order $L$ to solve the linear system according to eq. \ref{eq:linearsystemfinal} with preconditioning normalization of $\Theta = \frac{1}{\text{Tr}(A^*A)}$, which becomes;
 \begin{equation}
    \boldsymbol{x} = \prod_{j=0}^{i}\sum_{n=0}^{L}(I-\frac{A^*A}{\text{Tr}(A^*A)})^{n(L+1)^j}\frac{1}{\text{Tr}(A^*A)}A^*B,
     \label{eq:nnlinearsystem}
 \end{equation} 
 Where it is numerically seen, that even for random and badly conditioned matrices of $\text{cond}(A)=1\times10^6$, given an initial guess $\phi^{(0)} = I_{N\times N}$, the number of nests $i=37$ for an inception depth of $L=2$ is more than sufficient to have an accurate solution (see corollary 5). 
 \smallskip

\noindent \textbf{Corollary 4:}
The NI and CI can be simply derived from the NN by letting $L=1$ and $L=2$ respectively. Firstly, we directly get the NI from the NN of $i$ nests and $L=1$ depth;
$$
\phi_{L}^{(i+1)} = \sum_{n=0}^{1}(I-\phi_{L}^{(i)}\tilde{W})^n\phi_{L}^{(i)} = \phi^{(i+1)} = (2I - \phi^{(i)}\tilde{W})\phi^{(i)}.
$$
Similarly, from the NN of $i$ nests and inception depth of $L=2$, we find the CI, and by letting $L=3$ we find a $4^{th}$ order iterative method. This can be generalized, and used to show that the NN is the generalization of the $(L+1)^{th}$ order.
We can rewrite the NI and CI (from eq. \ref{eq:newtoniteration}, \ref{eq:cheby}) according to corollary (2), to find the following equations for the NI and CI respectively:
\begin{equation}
    Z_2^{(i+1)} = \sum_{n=0}^{2^{i+1}-1}(I-Z^{(0)}\tilde{W})^n Z^{(0)}, 
    \label{eq:newtoniteration1}
\end{equation}
\begin{equation}
    Z_3^{(i+1)} = \sum_{n=0}^{3^{i+1}-1}(I-Z^{(0)}\tilde{W})^n Z^{(0)},
    \label{eq:chebysheviteration}
\end{equation}
which shows that an $i^{th}$ order NI and CI is simply a NS of order $2^i - 1$ and $3^i -1$, respectively. This is an interesting artifact for matrix inversion in massive multiple input multiple output (mMIMO) systems \cite{Zhu2015}, as one of the commonly used methods is Chebyshev Neumann Series (CNS), where they perform the CI for $1,2$ iterations, so one maintains sparsity, and then this matrix is used as a preconditioner for the NS. This is indeed the most efficient method for CNS, as one should perform as many CI as possible whilst maintaining sparsity, and then using this as a preconditioner for the NS until convergence or maximum computational cost is reached. Therefore, with eq. \ref{eq:newtonfunction} and \ref{eq:chebysheviteration} we can give an analytical formula for the CNS method, displaying its empirical matrix inversion assessment accuracy, 
\begin{equation}
\begin{split}
    \tilde{W}^{-1} \approx  \\ \sum_{k = 0}^T (I-\sum_{n=0}^{3^{i+1}-1}(I-\phi^{(0)}\tilde{W})^n \phi^{(0)}\tilde{W})^k\sum_{n=0}^{3^{i+1}-1}(I-\phi^{(0)}\tilde{W})^n \phi^{(0)},
    \label{eq:CNS}
\end{split}
\end{equation}
where $\phi^{(i+1)} = \sum_{n=0}^{3^{i+1}-1}(I-\phi^{(0)}\tilde{W})^n \phi^{(0)}$ is computed per the CI in eq. \ref{eq:cheby} or \ref{eq:NNexplicit}, and $T$ is the number of terms included in the final NS. Therefore, the CNS yields a NS of order $S=3^{(i+1)T}$. The accuracy of the Newton-Neumann Series (NNS) can be derived following the same steps with $L=1$. Notably, the first summation can be substituted with the factorization shown in Corollary 6, to further reduce computational costs for higher $T\in \{2^s : s\in\mathbb{N}\}$.  
\smallskip

\noindent \textbf{Corollary 5:}
It has been shown that for the NI to convergence \cite{pan1991improved}, the maximum number of required iterations $i$ is such that,
\begin{equation}
    i = 2\log(\kappa(A)),
    \label{eq:kappa}
\end{equation}
where $\kappa(A) = ||A||_2||A^{-1}||_  2$ . Thus, the maximum number of required iterations can be generalized for any order by using Corollary 2 to find nests as a function of inception depth $i(L)$,
\begin{equation}
    i(L) = \log_{L+1}(2[\kappa(A)]^\frac{2}{\log_2(e)}) - 1.
    \label{eq:iasfunctionofL}
\end{equation} 
\smallskip

\noindent \textbf{Corollary 6:}
It can be shown that the NS can be factorized $\forall \gamma : \log_2(\gamma+1)\in\mathbb{N}$ (see corollary (1));
\begin{equation}
\sum_{n=0}^\gamma(I-\phi\tilde{W})^n\phi = \prod_{n=0}^{\log_2(\gamma+1)-1}(I+P^{2^n})\phi,
\label{eq:factorization}
\end{equation} 
where $P=I-\phi\tilde{W}$. This allows a generalization of the NS with a lower computational complexity than the traditional NS, for higher orders: $\gamma = 2^s -1, s\in\mathbb{N}$. The computational complexity of the factorization of the NN in eq. \ref{eq:factorization} is,
\begin{equation}
    C = 2\log_2(\gamma+1)N^3 + (\log_2(\gamma+1) + 1)N^2
    \label{eq:compcostfinal}
\end{equation}
for matrices that can be stored on the GPU. The computational complexity for massive sparse matrices, that cannot be stored as structured matrices on the GPU, is as stated below as shown in the \textit{appendix} (7),
\begin{equation}
    C = [\frac{(\log_2(\gamma+1))^2+\log_2(\gamma+1)-1}{2}]N^3 + [\log_2(\gamma+1)]^2N^2,
    \label{eq:factorizationcomplexity}
\end{equation}
assuming that the only matrix stored on the GPU is $P= I-\phi\tilde{W}$. This generalization is applicable to sparse matrices, as only a single sparse matrix is required to be stored, and for the application of the NN, allowing higher orders to be more competitive to the traditional NN. More specifically, for mMIMO systems, or other massive sparse linear systems, this could be applied to solve a linear system for order $\gamma -1$: $A\boldsymbol{x} = \boldsymbol{B}$, in the following manner;
\begin{equation}
    \boldsymbol{x} = \prod_{n=0}^{\log_2(\gamma+1)-1}(I+(I-\phi\tilde{W})^{2^n})\phi \boldsymbol{B},
\end{equation}
given that $\phi$ is a constant. This algorithm implementation is demonstrated in Algorithm 3.2, so that its applicability to sparse matrices is clear.

We can find a more suitable preconditioning matrix $\phi$ by doing a low order NS without loosing sparsity. For this, we can do one or a few CI to update our preconditioning matrix as seen in literature \cite{Zhu2015}, and then use $\phi^{(i)}$ from eq. \ref{eq:NNexplicit} with $L=2$ as the preconditioning matrix for the NS. We apply eq. \ref{eq:linearsystemfinal} to show that this product is simply a NS of order $3^i\gamma$, where the coefficient $3^i$ comes from the CI of iteration number $i$. Since the terms in this product commute, we can apply it to sparse matrices by storing the first variable as $\boldsymbol{B}_{temp} = (I + (I - \phi^{(1)})^L)\phi^{(1)}$, and then keep updating $\boldsymbol{B}_{temp}$ by multiplying it by the next term as seen in the pseudo-code section. For this, the computational complexity follows eq. \ref{eq:compcostfinal}, with $i=1$ and an added $2N^2$ from the CI with a diagonal preconditioner.

\section{Pseudo Code}
Here we will write a generalized code for the NN for any inception depth $L$, given the suggested preconditioners $\phi^{(0)} = I$ and $\Theta = \frac{1}{\text{Tr}(W)}$. 
\begin{algorithm}
\begin{algorithmic} 
\STATE W = import gpuArray(data);
\STATE i,L = x,y;
\STATE $\phi =$ Trace$(W)$; \%or eigenvalue theorem
\STATE $W_{tilde} = \phi W$;
\STATE $I =$ gpuArray$($eye$($dim$(W)))$;
\STATE $Z = I$;
\STATE \textbf{for} $j = 1:i$
\STATE $\quad$ $P = 1 - ZW_{tilde}$;
\STATE $\quad$ $S = I$;
\STATE $\quad$ \textbf{for} $l = 1:L$
\STATE $\quad\quad$ $S += P^l$;
\STATE $\quad$ \textbf{end for}
\STATE $\quad$ $Z = SZ$;
\STATE \textbf{end for}
\RETURN $Z\phi$
\end{algorithmic} 
\caption{The algorithm for the generalized NN, only applicable to structured matrices.}
\end{algorithm}

\begin{algorithm}
\caption{The algorithm for the generalized NN ($L+1=2$ shown), applicable to sparse linear systems.}
\begin{algorithmic}
\STATE W = import gpuArray(data); \%must be sparse
\STATE B = import gpuArray(data);
\STATE i,L = x,y;
\STATE $\phi =$ Trace$(W)$; \%or eigenvalue theorem
\STATE $W_{tilde} = \phi W$;
\STATE $I =$ gpuArray$($eye$($dim$(W)))$;
\STATE $Z = B$;
\STATE $P_{id} = 1 - ZW_{tilde}$;
\STATE \textbf{for} $k= 2. \wedge (0:(i-1))$
\STATE $\quad$ $Z = Z+ P_{id}^kZ$;
\STATE \textbf{end for}
\RETURN $Z\phi$

\end{algorithmic}
\end{algorithm}
Here, the inception depth and number of nests $(L,i)$ can be chosen as desired. This algorithm is embarassingly parallelizable on the GPU, and thus in light of Pan and Schreiber's attributions in 1991 \cite{pan1991improved}, provides a possibly fast algorithm for matrix inversion on powerful modern GPUs. Note, this is the simplest algorithm, and refer to the corollary's in order to update and optimize. Here, $Z\phi$ is the approximated inverse of the matrix. 

\section{Results}
This is a preliminary paper, which aims to provide the mathematical foundation of the NN and provide insight to some of the possible implications the NN can have. We show that any $\Theta$ that satisfies the normalization condition, causes convergence for the NN. We show an analytic explicit formula that relates the NS with the NN (thus NI and CI), which ultimately provides an analytic way to compare accuracy's of the different approximate matrix inversion algorithms. We also show different factorization forms of the NN, and how it can be applied to solve sparse linear systems. Furthermore, we demonstrate different computational costs, and provide different ways to alter these computational costs with a variety of parameters. Finally, we provide significant intuition into the different approximate matrix inversion methods, and demonstrate how they're all based on the NS.

\section{Conclusion}
This work investigates different approximate fast iterative matrix inversion methods. We developed a generalization of higher order methods: the Nested-Neumann, which in essence capitalizes on a suitable iterative update of the preconditioning matrix, through an inception depth $L$, and a given number of iterations $i$ (nests). Importantly, this can be explicitly written as a Neumann Series with order $(L+1)^i-1$, with a significantly less computational cost than the NS itself. We further explore methods to make the Nested Neumann of different orders applicable to sparse systems, and succeed as displayed by Algorithm 3.2 or Corollary 6. Interestingly, Corollary 1, also resulted in an interesting factorization of the series $\sum_{k=0}^L(x^k)$. We notice that the NN for orders 1 and 2, are respectively equivalent to Newton and Chebyshev iterations. We conclude that theoretically, the Chebyshev iteration always has a lower computational cost than the Newton iteration, however, requires an additional variable to be stored on GPU RAM $(I-\phi^{(i)}\tilde{W})$. 

This paper also mathematically investigated how the Nested Neumann uses this iterative update on the preconditioning matrix to turn itself into an series that has an order of convergence of $L+1$, by showing that it is essentially a Neumann Series with $(L+1)^{i+1}-1$ terms. It was also found that the optimal inception depth is $L=e-1$, which best approximates to $L=2$, which is the Chebyshev iteration. Through the factorization seen in Corollary 6, it is possible to make higher orders equivalently competitive with the Newton Iteration, however, still slightly less computationally efficient than the Chebyshev Iteration. Interestingly, the Nested Neumann provides a generalized higher order methodology, which can be expressed as a single Neumann Series. This has implications in diverse fields where massive multiple input multiple output (mMIMO) based linear systems need resolution, as the explicit accuracy for different order methods can be computed analytically by corollary (1,2), instead of experimentally comparing these different methodologies \cite{Zhu2015,Hashima2020,Albreem2021OverviewOP, qiang2020approximative}. \\

In line with the work by Pan and Schreiber in $1991$ \cite{pan1991improved} and new powerful parallel computers (GPUs), the Nested-Neumann, due to its high level of parallelizability, has potential to be a competitor to other fast iterative matrix inversion algorithms. We intend to further test its competitiveness against other matrix inversion algorithms or linear solvers with a GPU Cluster. We also demonstrate the ability to use the factorized Neumann Series to be applied to sparse matrices, with a lower computational cost than the traditional Neumann Series for higher orders than $3$. This could lead to more accurate sparse linear system resolution, which has particular importance in mMIMO systems \cite{Hashima2020}.

\section*{Acknowledgments}
This work has been performed in the framework of an internship supervised by Dr. Fares Mehouachi, audited by Sorbonne University of Abu Dhabi (SUAD), and sponsored by Technology Innovation Institute (TII). I would like to further thank Dr. Alejandro Tejedor (SUAD) for helpful and insightful discussions. 
\bibliographystyle{ieeetr} 
\bibliography{references}  

\appendix
\noindent \textbf{\large{1. Proposition:}} 

We define the preconditioning matrix $\phi_{L}^{(i+1)}$ iteratively, where $\phi^{(0)} = I_{N\times N}$:
\begin{equation}
  \label{eq:neumannnested}
  \phi_{L}^{(i+1)} = \sum_{n=0}^L (I - \phi^{(i)}\tilde{W})^n \phi^{(i)},
\end{equation}
where $\tilde{W} = \Theta W$ is normalized, such that $||I-\tilde{W}||_2 = || I - \Theta W ||_2 < 1$. Here, by similarity to the to the Neumann series, we see that for $i=0$:
\begin{equation}
  \lim_{L\to \infty} \phi_{L}^{(1)} = \tilde{W}^{-1},
\end{equation}
and thus, by multiplying both sides by $\Theta$, we can deduce $W^{-1}$, similar to eq. \ref{eq:Neumann}.

\begin{equation}
  W^{-1} \approx \phi_{L}^{(i+1)}\Theta = \sum_{n=0}^L (I - \phi^{(i)}\tilde{W})^n \phi^{(i)} \Theta
\end{equation}
Thus, by similarity to the Neumann Series, we see for $i=0$:
\begin{equation}
  \lim_{L\to \infty} \phi_{L}^{(1)}\Theta = W^{-1},
\end{equation}
which yields that for $i=0$, we have the simple Neumann Series as per eq. \ref{eq:Neumann}, where $\Theta\in\mathbb{R}$ or $\Theta\in\mathbb{M}_{N \times N}$ normalizes $W$ such that $|| I - \Theta W ||_2 < 1$. Now, as previously stated, the point of the Nested-Neumann series is to update the preconditioning matrix iteratively, and thus, we need to prove that the Nested-Neumann series converges to $\tilde{W}^{-1}$ as $i\to \infty, \forall L\in\mathbb{N}$, and as $L\to\infty, \forall i\in\mathbb{N}$, which implies we must show that:
\begin{eqnarray}
  \lim_{i\to\infty}\phi_{L}^{(i)} = \tilde{W}^{-1}, \forall L\in\mathbb{N}\setminus\{0\} \\
  \lim_{L\to\infty}\phi_{L}^{(i)} = \tilde{W}^{-1}, \forall i\in\mathbb{N}\setminus\{0\},
\end{eqnarray}
or, by Neumann convergence,
\begin{equation}
\label{eq:prove}
|| I - \phi_L^{(i)}\tilde{W} ||_2 < 1, \forall (i,L)\in\mathbb{N}\setminus\{0\}.
\end{equation}
Thus, if the Nested-Neumann Iteration converges to $\tilde{W}^{-1}$, we know the initial preconditioning term $\Theta$, such that we can find $W^{-1}$, the actual approximate inverse.\\

\noindent \textbf{\large{2. Proof:}} \textbf{Nested-Neumann Convergence}
We start the proof by assuming that $|| I - \phi^{(i)}\tilde{W} ||_2 < 1$, and show that $ || I - \phi_{L}^{(i+1)}\tilde{W}||_2 < 1$, $\forall L\in\mathbb{N}$ by mathematical induction. \\
\fbox{$L=0$}
$$
\phi_{L=0}^{(i+1)} = \sum_{n=0}^0 (I - \phi^{(i)}W)^n \phi^{(i)} \implies \phi_{L=0}^{(i+1)} = \phi^{(i)} ,
$$
thus,
$$
|| I - \phi_{L=0}^{(i+1)}\tilde{W} ||_2 = || I - \phi^{(i)}\tilde{W} ||_2 < 1 \quad \checkmark
$$
by initial assumption. \\
\fbox{$L=1$} 
$$
\phi_{L=1}^{(i+1)} = \sum_{n=0}^1 (I - \phi^{(i)}\tilde{W})^n \phi^{(i)}$$ 
which implies that,
$$ \phi_{L=1}^{(i+1)} = (2I - \phi^{(i)}\tilde{W}) \phi^{(i)} ,
$$
thus, for the norm:
\begin{align*}
|| I - \phi_{L=0}^{(i+1)}\tilde{W} ||_2 = || I - 2\phi^{(i)}\tilde{W} + (\phi^{(i)}W)^2 ||_2 \\ = ||(I - \phi^{(i)}\tilde{W})^2||_2 
\end{align*}
which implies that,
$$
|| I - \phi_{L=0}^{(i+1)}\tilde{W} ||_2 \leq || I - \phi^{(i)}\tilde{W} ||_2 \cdot || I - \phi^{(i)}\tilde{W} ||_2 < 1, \quad \checkmark
$$
which is true by initial assumption. \\
Assume true for \fbox{$L=k$}, which implies, by similarity to $L=1$:
$$
|| (I - \phi^{(i)}\tilde{W})^{k+1} ||_2 < 1.
$$
\fbox{$L = k + 1$} 
$$
\phi_{L=k+1}^{(i+1)} = \sum_{n=0}^{k+1} (I - \phi^{(i)}\tilde{W})^n \phi^{(i)},
$$
thus, for the norm: 
\begin{align*}
|| I - \phi_{L = k+1}^{(i+1)}\tilde{W}||_2 = || I - \sum_{n=0}^{k+1} (I - \phi^{(i)}\tilde{W})^n \phi^{(i)}\tilde{W} ||_2 \\ = || (I - \phi^{(i)}\tilde{W})^{k+2} ||_2 
\end{align*}
which implies that,
$$
|| I - \phi_{L = k+1}^{(i+1)}\tilde{W}||_2 \leq || (I - \phi^{(i)}\tilde{W})^{k+1} ||_2 \cdot || I - \phi^{(i)}\tilde{W} ||_2 < 1, \quad \checkmark
$$
by initial assumption and $L = k$ assumption. Thus the proof is done, and it has been shown:
\begin{equation}
  \forall L\in\mathbb{N}, \quad || I - \phi_{L}^{(i+1)}\tilde{W} ||_2 < 1, \quad \text{given} \quad || I - \phi_{L}^{(i)}\tilde{W} ||_2 < 1.
\end{equation}
\\
Now, we will start the second part of the proof: proving that the Nested-Neumann holds $\forall i\in\mathbb{N}$.
We start the proof by choosing $\phi^{(0)}: || I - \phi^{(0)}\tilde{W} ||_2 < 1$ (ex. $\phi^{(0)} = I$), and show that $ || I - \phi_{L}^{(i)}\tilde{W}||_2 < 1$, $\forall i\in\mathbb{N}$ by mathematical induction. \\
\fbox{$i=0$} is trivial, as $\phi^{(0)}$ is chosen such that:
$$
|| I - \phi^{(0)}\tilde{W} ||_2 < 1, \quad \checkmark
$$
\fbox{$i=1$}
$$
\phi_{L}^{(1)} = \sum_{n=0}^{L} (I -\phi^{(0)}\tilde{W})^n\phi^{(0)},
$$
for the norm:
\begin{align*}
|| I - \phi_{L}^{(1)}\tilde{W} ||_2 = || (I -\sum_{n=0}^{L}(I-\phi^{(0)}\tilde{W})^n\phi^{(0)}\tilde{W} ||_2 \\= ||(I-\phi^{(0)}\tilde{W})^{L+1} ||_2
\end{align*}
which implies,
$$
||I - \phi_{L}^{(1)}\tilde{W} ||_2 \leq \prod_{n=0}^{L+1}|| I - \phi^{(0)}\tilde{W}||_2 < 1, \quad \checkmark
$$
as $|| I - \phi^{(0)}\tilde{W}||_2 < 1$ is a precondition, and $\phi^{(0)}$ is chosen to fulfill that statement. \\
Now, we assume \fbox{i = k} to be true, which implies:
$$
||I - \phi_{L}^{(k)}\tilde{W} ||_2 < 1.
$$
\fbox{$i = k+1$} 
$$
\phi_{L}^{(k+1)} = \sum_{n=0}^L (I - \phi_{L}^{(k)}\tilde{W})^n \phi_{L}^{(k)},
$$
for the norm,
\begin{align*}
|| I - \phi_{L}^{(k+1)} \tilde{W}||_2 = || I - \sum_{n=0}^L (I - \phi_{L}^{(k)}\tilde{W})^n \phi_{L}^{(k)} \tilde{W}||_2 \\= || (I - \phi_{L}^{(k)}\tilde{W})^{L+1} ||_2
\end{align*}
which finally implies that,
$$
 || I - \phi_{L}^{(k+1)} \tilde{W}||_2 \leq \prod_{n=0}^{L+1}|| I - \phi^{(k)}\tilde{W}||_2 < 1, \quad \checkmark
$$
which is true by by $i = k$ assumption. Thus the proof is done, and it has been shown:
\begin{equation}
  \forall i\in\mathbb{N}, \quad || I - \phi_{L}^{(i+1)}\tilde{W} ||_2 < 1, \quad \text{if} \quad || I - \phi_{L}^{(i)}\tilde{W} ||_2 < 1.
\end{equation}

Therefore, as the norm $||I - \phi_L^{(i)}\tilde{W}||<1, \forall i,L\in\mathbb{N}$, we have proven by Neumann convergence that:

\begin{align*}
  \lim_{i\to\infty}\phi_{L}^{(i)} = \tilde{W}^{-1}, \forall L\in\mathbb{N}\setminus\{0\} \\
  \lim_{L\to\infty}\phi_{L}^{(i)} = \tilde{W}^{-1}, \forall i\in\mathbb{N}\setminus\{0\}.
\end{align*} \\

\noindent \textbf{\large{3. Corollary:}} \textbf{Explicit Nested-Neumann}\\

We attempt to rewrite the Nested-Neumann such that we can find the $(i+1)^{th}$ iteration as a function of $i,L, \phi^{(0)}$. The Nested Neumann can be written as per eq. \ref{eq:nestedneumann}. 
$$
\phi_{L}^{(i+1)} = \sum_{n=0}^{L}(I-\phi_{L}^{(i)}\tilde{W})^n\phi_{L}^{(i)},
$$
which implies that $\phi^{(i)}$ can be written as follows;
\begin{equation}
\phi_{L}^{(i)} = \sum_{n=0}^{L}(I-\phi_{L}^{(i-1)}\tilde{W})^n\phi_{L}^{(i-1)}.
\label{eq:neumannminus1}
\end{equation}
Now substituting eq. \ref{eq:neumannminus1} into eq.\ref{eq:nestedneumann}, yields the following.
$$
\phi_{L}^{(i+1)} = \sum_{n=0}^{L}(I-\sum_{p=0}^{L}(I-\phi_{L}^{(i-1)}\tilde{W})^p\phi_{L}^{(i-1)} \tilde{W})^n \phi^{(i)},
$$
which can be simplified as previously,
$$
\phi_{L}^{(i+1)} = \sum_{n=0}^{L}(I-\phi^{(i-1)} \tilde{W})^{n(L+1)} \phi^{(i)},
$$
doing this same operation iteratively until $\phi^{(0)}$ yields:
\begin{equation}
\phi_{L}^{(i+1)} = \sum_{n=0}^{L}(I-\phi^{(0)} \tilde{W})^{n(L+1)^i} \phi^{(i)}
\label{eq:phifromphi0}
\end{equation}
Now doing the same for the $\phi^{(i)}$ term on the outside:
$$
\phi_{L}^{(i+1)} = \sum_{n=0}^{L}(I-\phi^{(0)} \tilde{W})^{n(L+1)^i} \sum_{n=0}^{L}(I-\phi_{L}^{(i-1)}\tilde{W})^n\phi_{L}^{(i-1)},
$$
now simply using eq. \ref{eq:phifromphi0} on the outside term and simplifying,
$$
\phi_{L}^{(i+1)} = \sum_{n=0}^{L}(I-\phi^{(0)} \tilde{W})^{n(L+1)^i} \sum_{n=0}^{L}(I-\phi^{(0)} \tilde{W})^{n(L+1)^{i-1}} \phi_{L}^{(i-1)},
$$
which done iteratively until $\phi^{(0)}$ on the outside will yield the final equation:
\begin{equation}
    \phi_L^{(i+1)} =  \prod_{j=0}^{i}\sum_{n=0}^{L}(I-\phi^{(0)} \tilde{W})^{n(L+1)^j}\phi^{(0)}
    \label{eq:nestedneumannproductsum}
\end{equation} \\

\noindent \textbf{\large{4. Corollary}}: \textbf{Nested-Neumann as a Neumann Series}\\ 

Now, we will rewrite the Nested-Neumann as a Neumann series only, by using geometric series to remove the sequential product term. The Neumann Component (sum) of eq. \ref{eq:nestedneumannproductsum} can be rewritten as a geometric series:
\begin{equation}
    \sum_{n=0}^{L}(I-\phi^{(0)} \tilde{W})^{n(L+1)^j}\phi^{(0)} = \sum_{n=0}^L q^n = \frac{1-q^{L+1}}{1-q}\phi^{(0)},
    \label{eq:geometricseries}
\end{equation}
where $q = (I-\phi^{(0)}\tilde{W})^{(L+1)^j}$.
Therefore, substituting this into eq. \ref{eq:nestedneumannproductsum} yields,
$$
 \phi_L^{(i+1)} =  \prod_{j=0}^{i} \frac{1-(I-\phi^{(0)}\tilde{W})^{(L+1)^j(L+1)}}{1-(I-\phi^{(0)}\tilde{W})^{(L+1)^j}} \phi^{(0)},
$$
which is trivially equal to:
$$
\phi_L^{(i+1)} =  \prod_{j=0}^{i} \frac{1-(I-\phi^{(0)}\tilde{W})^{(L+1)^{j+1}}}{1-(I-\phi^{(0)}\tilde{W})^{(L+1)^j}} \phi^{(0)},
$$
Now opening up the product term and canceling out terms yields:
$$
\phi^{(i+1)} = \frac{I - (I - \phi^{(0)}\tilde{W})^{(L+1)^{j+1}}}{I - (I-\phi^{(0)}\tilde{W})} \phi^{(0)},
$$
which is the result of the geometric series in eq. \ref{eq:geometricseriesneumann},
\begin{equation}
    \phi^{(i+1)} = \sum_{n=0}^{(L+1)^{i+1}-1}(I-\phi^{(0)}\tilde{W})^n \phi^{(0)}.
    \label{eq:geometricseriesneumann}
\end{equation}
This equation above shows that the Nested-Neumann is simply a Neumann series of an order scaling with $(L+1)^{i+1}-1$. This implies, that an enormous amount of products can be computed in very few iterations. For example: $2^{50} = 1.12\times 10^{50}$ iterations, which would normally require $1.12\times 10^{50}$ matrix multiplications and $1.12\times 10^{50}$ matrix additions, whereas with the Nested-Neumann, this can be computed in $2\times50 = 100$ matrix multiplications and $2\times 50$ matrix additions. \\

\noindent \textbf{\large{5. Proof}}: \textbf{Rate of Convergence}\\ 
 \textit{Definition:} A sequence $X_n$ that converges to $r$ is said to have order of convergence $\alpha \geq 1$ and rate of convergence $\mu$ if for any norm, \cite{schatzman2002numerical}
\begin{equation}
    \lim_{n\to\infty} \frac{|| x_{n+1} -r ||}{|| x_n - r||^\alpha} = \mu.
    \label{eq:convergence}
\end{equation}
We know that $\phi^{(i)}_L$ is a sequence such that:
$$
\lim_{i\to\infty}\phi_L^{(i)} = \tilde{W}^{-1},
$$
now we  define the residual ($x_i$) of the Nested-Neumann: $x_i = \phi_L^{(i)} - \phi_L^{(i-1)}$, which can be rewritten via. eq. \ref{eq:geometricseriesneumann};
$$
x_i = \sum_{n=(L+1)^{i-1}}^{(L+1)^{i}-1}(I-\phi^{(0)}\tilde{W})^n \phi^{(0)},
$$
where $\lim_{i\to\infty}x_i = 0$, and thus substituting into eq. \ref{eq:convergence} yields,
\begin{equation}
   \lim_{i\to\infty} \frac{||\sum_{n=(L+1)^{i}}^{(L+1)^{i+1}-1}(I-\phi^{(0)}\tilde{W})^n \phi^{(0)}||}{||\sum_{n=(L+1)^{i-1}}^{(L+1)^{i}-1}(I-\phi^{(0)}\tilde{W})^n \phi^{(0)}||^\alpha} = \mu.
   \label{eq:stage1}
\end{equation}
Now, the product of norms is always larger than the norm of a product, and thus we can write,
\begin{equation}
\begin{split}
    ||\sum_{n=(L+1)^{i-1}}^{(L+1)^{i}-1}(I-\phi^{(0)}\tilde{W})^n \phi^{(0)} || \geq \\ ||[\sum_{n=(L+1)^{i-1}}^{(L+1)^{i}-1}(I-\phi^{(0)}\tilde{W})^n \phi^{(0)}]^\alpha ||.
\end{split}
    \label{eq:norminequality}
\end{equation}
Therefore, we can turn eq. \ref{eq:stage1} into an inequality by using eq. \ref{eq:norminequality},
\begin{equation}
\mu \leq \lim_{i\to\infty} \frac{||\sum_{n=(L+1)^{i}}^{(L+1)^{i+1}-1}(I-\phi^{(0)}\tilde{W})^n \phi^{(0)}||}{||[\sum_{n=(L+1)^{i-1}}^{(L+1)^{i}-1}(I-\phi^{(0)}\tilde{W})^n \phi^{(0)}]^\alpha||},
\label{eq:stage2}
\end{equation}
and as the matrices are positive semi-definite, we can make the following statement:
\begin{equation}
\begin{split}
    ||[\sum_{n=(L+1)^{i-1}}^{(L+1)^i-1}(I-\phi^{(0)}\tilde{W})^n \phi^{(0)}]^{\alpha} || \geq \\ ||\sum_{n=(L+1)^{i-1}}^{(L+1)^i-1}(I-\phi^{(0)}\tilde{W})^{n\alpha} (\phi^{(0)})^\alpha||
    \label{eq:normpsd}
\end{split}
\end{equation}
now using eq. \ref{eq:normpsd} to create a further inequality for eq. \ref{eq:stage2} yields,
$$
\mu \leq \lim_{i\to\infty} \frac{||\sum_{n=(L+1)^{i}}^{(L+1)^{i+1}-1}(I-\phi^{(0)}\tilde{W})^n \phi^{(0)}||}{||\sum_{n=(L+1)^{i-1}}^{(L+1)^{i}-1}(I-\phi^{(0)}\tilde{W})^{n\alpha} (\phi^{(0)})^\alpha||} = \mu_{sup},
$$
where $\mu_{sup}$ is the suprememum that the $\mu$ value is bounded by. Therefore, if we can prove that for a given $\alpha$, there exists a finite $\mu_{sup}$, the proof is done. Let $\alpha = L + 1$ by hypothesis, then by changing the indices accordingly, we find, 
\begin{equation}
    \mu \leq \lim_{i\to\infty} \frac{||\sum_{n=(L+1)^{i}}^{(L+1)^{i+1}-1}(I-\phi^{(0)}\tilde{W})^n \phi^{(0)}||}{||\sum_{n=(L+1)^{i}}^{(L+1)^{i+1}-1}(I-\phi^{(0)}\tilde{W})^{n} [\phi^{(0)}]^{L+1}||} = \mu_{sup},
\end{equation}
which by simplification yields the solution for the supremum of $\mu$,
\begin{equation}
    \mu \leq \mu_{sup} = \frac{1}{||\phi^{(0)}||^L}
    \label{eq:musup}
\end{equation}
which, is always a finite number, and specifically, if $\phi^{(0)} = I$, then we have that this is uniquely equal to $1$ using the spectral norm. Thus, we have proved that the rate of convergence is of order $L+1$. \\

\noindent \textbf{\large{6. Proof}}: \textbf{Optimal Inception Depth}\\ 
The computational cost of the Nested Neumann ($C$) is:
\begin{equation}
    C = i(L+1)N^3 + i(L+1)N^2.
    \label{eq:cost}
\end{equation}
The accuracy of the Nested Neumann can be seen as per the Neumann Series described in eq. \ref{eq:geometricseriesneumann}, the last order of the Neumann Series ($\epsilon$) can be described as,
\begin{equation}
    \Gamma = (L+1)^{i+1} - 1 
    \label{eq:epsilon}
\end{equation}
Now, for large matrices, $N^3 >> N^2$, and thus, we can approximate the computational cost to be $C = i(L+1)N^3$. Consider we can afford $KN^3, K\in\mathbb{N}$ operations,
$$
KN^3 = i(L+1)N^3 \implies i = \frac{k}{L+1},
$$
 we seek to maximize $\epsilon$ for this given $K$, 
$$
\Gamma = (L+1)^{\frac{K}{L+1} + 1} - 1.
$$
Therefore, to find max, we take the derivative with respect to the optimization parameter $L$, the inception depth, and find, 
\begin{equation}
    \frac{\partial \Gamma}{\partial L} = \frac{\partial}{\partial L} [(L+1)^{\frac{K}{L+1} + 1} - 1] = K(L+1)^{\frac{K}{L+1}-1}(1-log_2(L+1)),
\end{equation}
which implies that,
\begin{equation}
\frac{\partial \Gamma}{\partial L} = 0 \iff L = e-1
\label{eq:optimum}
\end{equation}
Thus, per eq. \ref{eq:optimum} we have shown that the optimal depth of inception is $L=e-1$, however, $L\in\mathbb{N}$, and thus $L=1,2$ are the optimum, with $L=2$ being slightly superior to $L=1$. However, with $L=2$, there is an additional variable that needs to be stored $(I - \phi^{(i)}\tilde{W})$, which can be avoided through $L=1$. Therefore, as a function of the GPU specs, either the Newton or Chebyshev Iterations can be the optimum. However, theoretically speaking, if storing data was free, the Chebyshev Iteration is the most efficient.   \\

\noindent \textbf{\large{7. Proof}}: \textbf{NS Factorization}\\ 
The $L^{th}$ order NS can be written as per eq.\ref{eq:Neumann},
$$
  W^{-1} \approx \sum_{n=0}^{L} [(I - \phi \tilde{W})^n] \phi = \sum_{n=0}^{L}P^n, \quad ||I - \phi W ||_2 < 1 ,
  $$
where $P=I-\phi\tilde{W}$. Now, assume that $L: \log_2{L+1}\in\mathbb{N}$, then $x^L$ is the highest order, and consequently, the term $x^{\frac{L+1}{2}}$ can be factored out of the largest half of the terms, assuming that $L>1$, and satisfies the previous condition:
\begin{equation}
    \sum_{n=0}^L P^n = (1+P^{\frac{L+1}{2}})\sum_{n=0}^{L-\frac{L+1}{2}}P^n,
    \label{eq:x^Lassmallerterms}
\end{equation}
which can be done iteratively, until the upper bound in the remaining summation is $1$, at which point, the equation will simply become as the one in eq. \ref{eq:factorization},
\begin{equation}
\sum_{n=0}^L(1-\phi\tilde{W})^n\phi = \prod_{n=0}^{\log_2(L+1)-1}(1+P^{2^n})\phi,
\label{eq:cor7final}
\end{equation}
where as previously stated: $P=I-\phi\tilde{W}$. This factorization is an important finding as it does not require any non-sparse matrices to be stored on the GPU, and thus allows application to sparse matrices. \\

\noindent \textbf{\large{8. Demonstration}}: \textbf{Computational Complexity of Factorized NS for Massive Sparse Matrices}\\ 
The computational complexity of matrix powers is a function of the specific power. Deriving a matrix power is divided into two subcategories, squaring and doubling. Through this method, only the initial matrix needs to be stored on the RAM, and any matrix power can be returned. For example, for a matrix $A^{14}$, one would do the following operations:
$$
A^1 \to A^2:\text{squaring}
$$
$$
A^2 \to A^3: \text{doubling}
$$
$$
A^3 \to A^6: \text{squaring}
$$
$$
A^6 \to A^7: \text{doubling}
$$
$$
A^7 \to A^{14}: \text{squaring}
$$
and thus $A^{14}$ would be found using minimal operations. Fortunately, from eq. \ref{eq:cor7final}, we see that the necessary powers are all power square multiples, that means $1,2,4,8,16,32,64...$. This means, that the computational complexity of these would follow $N^3, N^3, 2N^3, 3N^3, 4N^3, 5N^3, 6N^3$, as the first variables would be a multiplication to store $P = I - \phi\tilde{W}$, and $P^2$ would be 1 multiplication, $P\times P$. The rest, would come from simply doing squaring operations on $P$. Therefore, we notice for the computational complexity, that we have an arithmetic sequence that needs to have all of its terms multiplied, with $u_1 = 1$, and $u_f = \log_2(\gamma+1)$ (where $\gamma$ is the order of the NS), with a difference $d=1$, and an extra two multiplications from storing the matrix $P$, and from multiplying by $\phi$, and thus we can define the computational cost:
\begin{equation}
 C = 2\log_2(\gamma+1)N^3 + (\log_2(\gamma+1) + 1)N^2
\end{equation}

\end{document}